# Flexoelectric Polarization Enhancement in Paraelectric $BaHfO_3$ via Strain Gradient Engineering

*Timo Piecuch[1,2], Nina Daneu[3], Jeffrey A. Brock[1,4], Xiaochun Huang[1,2], Tina Radoševič[3], Arnold M. Müller[5], Christof Vockenhuber[5], Christof W. Schneider[1], Thomas Lippert[1,2], Nick A. Shepelin[1]**

[1]Center for Neutron and Muon Sciences, Paul Scherrer Institute, 5232 Villigen PSI, Switzerland

[2]Laboratory of Inorganic Chemistry, Department of Chemistry and Applied Biosciences, ETH Zürich, 8093 Zurich, Switzerland

[3]Advanced Materials Department, Jožef Stefan Institute, 1000 Ljubljana, Slovenia

[4]Laboratory for Mesoscopic Systems, Department of Materials, ETH Zürich, 8093 Zurich, Switzerland

[5]Laboratory of Ion Beam Physics, ETH Zürich, 8093, Zürich, Switzerland

**Keywords:** $BaHfO_3$, paraelectric, dielectric, strain gradient, flexoelectricity, thin film, perovskite oxide

## Abstract

Flexoelectricity - polarization induced by strain gradients - offers a route to polar functionality in centrosymmetric dielectrics, where traditional piezoelectric effects are absent. This study investigates the flexoelectric effect in epitaxial $BaHfO_3$ (BHO) thin films, a centrosymmetric and paraelectric perovskite. While a large lattice mismatch induces defect-driven relaxation, a coherently grown BHO film undergoes elastic relaxation, forming intrinsic strain gradients exceeding $10^5$ $m^{-1}$. A 29-fold enhancement in spontaneous polarization is observed at an electric field of 4 MV $cm^{-1}$ for BHO exhibiting a strain gradient compared to relaxed BHO. This enhancement is attributed to flexoelectric coupling, which is isolated from ferroelectric and piezoelectric contributions due to the centrosymmetric nature and the absence of phase transitions in BHO. The findings establish a clear link between engineered strain gradients and enhanced polarizability in oxide thin films, offering a benchmark system for deconvoluting the flexoelectric effect from other polar effects. These results provide a basis for exploiting flexoelectricity in dielectric devices and advance the fundamental understanding of strain-coupled phenomena in functional oxides.

## Introduction

Strain gradients play a fundamental role in determining the structural distortions that give rise to polar order, particularly at the nanoscale. In crystalline solids, strain gradients describe spatially inhomogeneous variations in strain, which can significantly influence physical properties such as polarization, defect dynamics, and phase stability.[1-4] Understanding and controlling these gradients is crucial for designing next-generation functional materials with tailored polar responses.[5,6]

In epitaxial thin films, the misfit strain is primarily governed by the lattice mismatch between the film and its underlying substrate or buffer layer, given by $\varepsilon=(a_f-a_s)/a_s$, where $a_f$ and $a_s$ correspond to the in-plane lattice parameters of the film and the substrate, respectively.[7,8] In oxides, a lattice misfit strain above 1-2% typically leads to strain relaxation at the interface via the formation of structural defects (e.g., point defects and threading dislocations), driving the material toward its bulk lattice parameter **(Figure 1(a))**. Conversely, when the misfit strain is sufficiently small and the growth parameters are sufficiently optimized, the film remains coherently strained to the underlying lattice of the substrate. A coherent growth results in a gradual relaxation throughout the structure **(Figure 1(b))**, governed by the mechanical properties of the film. This strain evolution can introduce substantial strain gradients across the film thickness. [9-11]

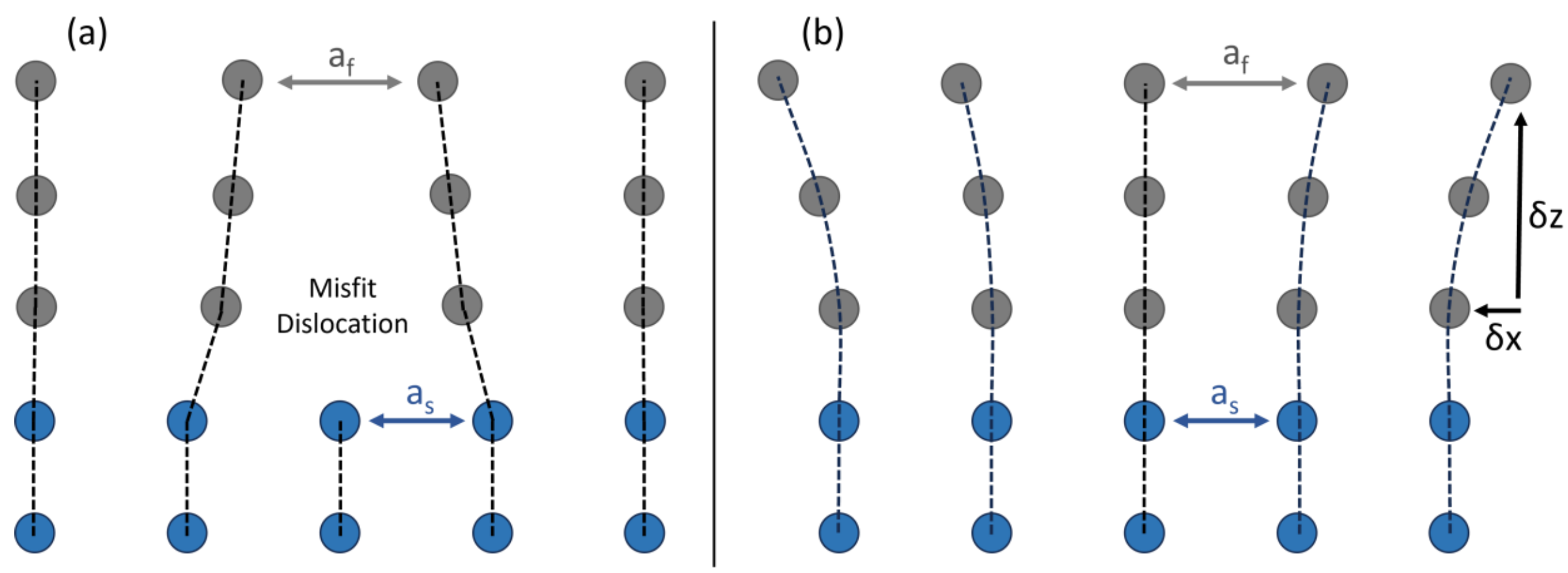

**Figure 1.** Schematic of a film (grey) grown on a substrate or buffer layer (blue), with lattice constants $a_f$ and $a_s$ respectively. In case of a large lattice mismatch (a), relaxation is mediated by the formation of misfit dislocations, while in the epitaxial case (b), strain relaxation occurs via a continuous strain gradient.

An intriguing consequence of strain gradients in dielectric materials is the flexoelectric effect, a phenomenon in which a spatial variation in strain ($\nabla\varepsilon$) induces an electric polarization P that manifests as bound surface charge.[1–4] The general constitutive relation for the direct flexoelectric effect can be expressed as $P_i = \mu_{ijkl}\frac{\partial \varepsilon_{jk}}{\partial x_l}$ (1) where $\mu_{ijkl}$ is the fourth-rank flexoelectric tensor (units: C $m^{-1}$), $\varepsilon_{jk}$ is the strain tensor, and $\partial\varepsilon_{jk} / \partial x_l$ is the strain gradient. Here, *i, j, k,* and *l* are tensor indices referring to the Cartesian coordinates of the polarization component (*i*), the strain tensor components (*j, k*), and the spatial direction of the strain gradient (*l*). This relation shows that the induced polarization is proportional to both the magnitude of the strain gradient and the intrinsic flexoelectric coefficient. Unlike piezoelectricity, which is symmetry-restricted to non-centrosymmetric space groups and is an intrinsic property fixed by the crystal chemistry, flexoelectricity is symmetry-allowed in all dielectrics, including centrosymmetric crystals.[12–16] This makes it particularly attractive for systems where ferroelectric or piezoelectric materials are unsuitable due to chemical, thermal, or structural incompatibilities. Potential applications include nanoscale energy harvesters, tunable capacitors, and electromechanical sensors. Nevertheless, despite its fundamental importance, flexoelectricity has been largely overlooked due to its typically small magnitude in bulk materials.[1,5,6]

The potential of flexoelectricity is enhanced in epitaxial thin films, since elastic strain relaxation results in much larger strain gradients (up to $10^6$ $m^{-1}$ in this work) compared to relaxation over bulk scale (up to $10^{-1}$ $m^{-1}$ in $SrTiO_3$ (STO) single crystals).[1,17,18] An additional advantage of thin films is the precise control of the initial strain state, which is determined by the lattice mismatch between the film and the substrate.[7,8] The unique mechanical properties of perovskite oxides enable tunable strain gradients through elastic relaxation mechanisms. In a thin-film geometry, such strain gradients often have a strong out-of-plane component (z) due to lattice-parameter relaxation across the film thickness. This geometry naturally favors an out-of-plane polarization component, where the polarization vector is aligned along the film normal and can be directly probed in vertical capacitor structures. The magnitude of this polarization is determined by the flexoelectric tensor components $\mu_{zzzz}$ and $\mu_{zjjz}$ ($j \neq z$). Because strain gradients in elastically relaxed epitaxial films typically scale as $\nabla\varepsilon \propto 1/t$ (where $t$ is the film thickness), the flexoelectric polarization is expected to follow approximately $P \propto \mu/t$ in the absence of other enhancement mechanisms. Strictly speaking, however, the relaxation rate is governed by the mismatch strain and the elastic compliance ($\propto$1/biaxial modulus), such that the gradient remains nearly constant until the film approaches its critical relaxation thickness. This inverse thickness dependence means that thinner films - within the elastic relaxation regime - can exhibit significantly larger flexoelectric polarization than thicker ones.[1,19] The ability to engineer and manipulate these strain fields in thin films opens new avenues for tuning material functionalities, particularly in electromechanical and dielectric applications.[20,21] For instance, research on tensile-strained $HoMnO_3$ films has revealed strain gradients several orders of magnitude larger than those in bulk oxides, leading to significant flexoelectric responses.[22,23]

$BaHfO_3$ (BHO) emerges as an ideal model system for studying flexoelectricity in epitaxial oxide thin films due to its unique combination of properties. As a centrosymmetric perovskite crystallizing in the cubic *Pm-3m* space group, BHO eliminates the interference of linear and nonlinear dielectric effects, allowing for a clear distinction of flexoelectric contributions. The paraelectric nature of this material, with no known phase transitions, ensures stability across a broad range of temperatures, strain states and electric fields.[24] While STO is widely used as a benchmark for flexoelectric research, its incipient ferroelectricity and temperature-dependent dielectric constant can complicate the isolation of pure flexoelectric contributions. Previous studies have demonstrated epitaxial strain-induced ferroelectric phase transitions in STO.[25,26] In contrast, BHO retains robust paraelectric behavior with a stable dielectric response, offering a more robust platform for quantitative analysis. Additionally, the atomic planes in BHO (i.e., $Ba^{2+}O^{2-}$ and $Hf^{4+}O_2^{2-}$) exhibit balanced charge distributions, eliminating polarization arising from charged plane contributions.[27] The material also exhibits a high dielectric constant, which enhances its potential for dielectric applications.[28] These properties, along with a pronounced mechanical stability, make it well-suited for integration into advanced electronic devices.[24,29]

To achieve a gradual strain relaxation throughout the film, the BHO films have been grown on STO (001) substrates buffered with epitaxially-grown La-doped $BaSnO_3$ (LBSO, 7% La) that acted as the bottom electrode and set the interfacial strain to -1.2%. Symmetric capacitor structures have been prepared for electrical measurements. These results are compared to a BHO capacitor grown on a conductive $SrRuO_3$ (SRO) buffer layer, corresponding to an interfacial lattice mismatch of -6.1%, which results in a misfit dislocation-driven relaxation at the SRO/BHO interface to release the strain. The strain states and strain gradients have been measured by high-resolution X-ray diffraction (HRXRD), and scanning transmission electron microscopy (STEM) has been employed to locally verify the sharp interfaces with no signs of atomic diffusion between the layers. These analyses reveal pronounced strain gradients as the dominant relaxation mechanism in capacitors utilizing LBSO electrodes. To elucidate the polar behavior of both systems, temperature-dependent dielectric spectroscopy and polarization-electric field hysteresis measurements have been conducted. A striking 29-fold increase in spontaneous polarization was observed in comparison to capacitor structures exhibiting dislocation-mediated relaxation on an SRO bottom electrode, revealing the impact of strain gradient-driven flexoelectric effects.

This research provides a fundamental basis for achieving pronounced strain gradients, along with significantly enhanced field-dependent polarization, in high-quality epitaxial BHO thin films. For the first time, such an investigation is conducted on a paraelectric, centrosymmetric material with no known phase transitions, offering a pathway to isolate and study pure flexoelectric effects without interference from ferroelectric or piezoelectric contributions.

## Results and Discussion

Symmetric capacitor structures were fabricated on STO substrates using two different bottom electrode materials. Each layer was grown *via* pulsed laser deposition (PLD), without breaking vacuum, under previously iteratively optimized process conditions. As a reference, BHO was deposited on a conductive SRO buffer layer, corresponding to a large interfacial lattice mismatch of -6.1% (**Figure 2a**). In contrast, in the second configuration, BHO was epitaxially grown on LBSO, establishing an interfacial strain of -1.2% between the bottom electrode and the dielectric layer (**Figure 2b**). The top and bottom electrode layers were grown to a thickness of 40 nm, while the BHO was grown to a thickness of 100 nm. Reciprocal space mapping (RSM) analysis was conducted to elucidate the dominant strain relaxation mechanism in each structure, which formed the basis for the electrical characterization of the two systems (**Figure 2c** and **Figure 2d**).

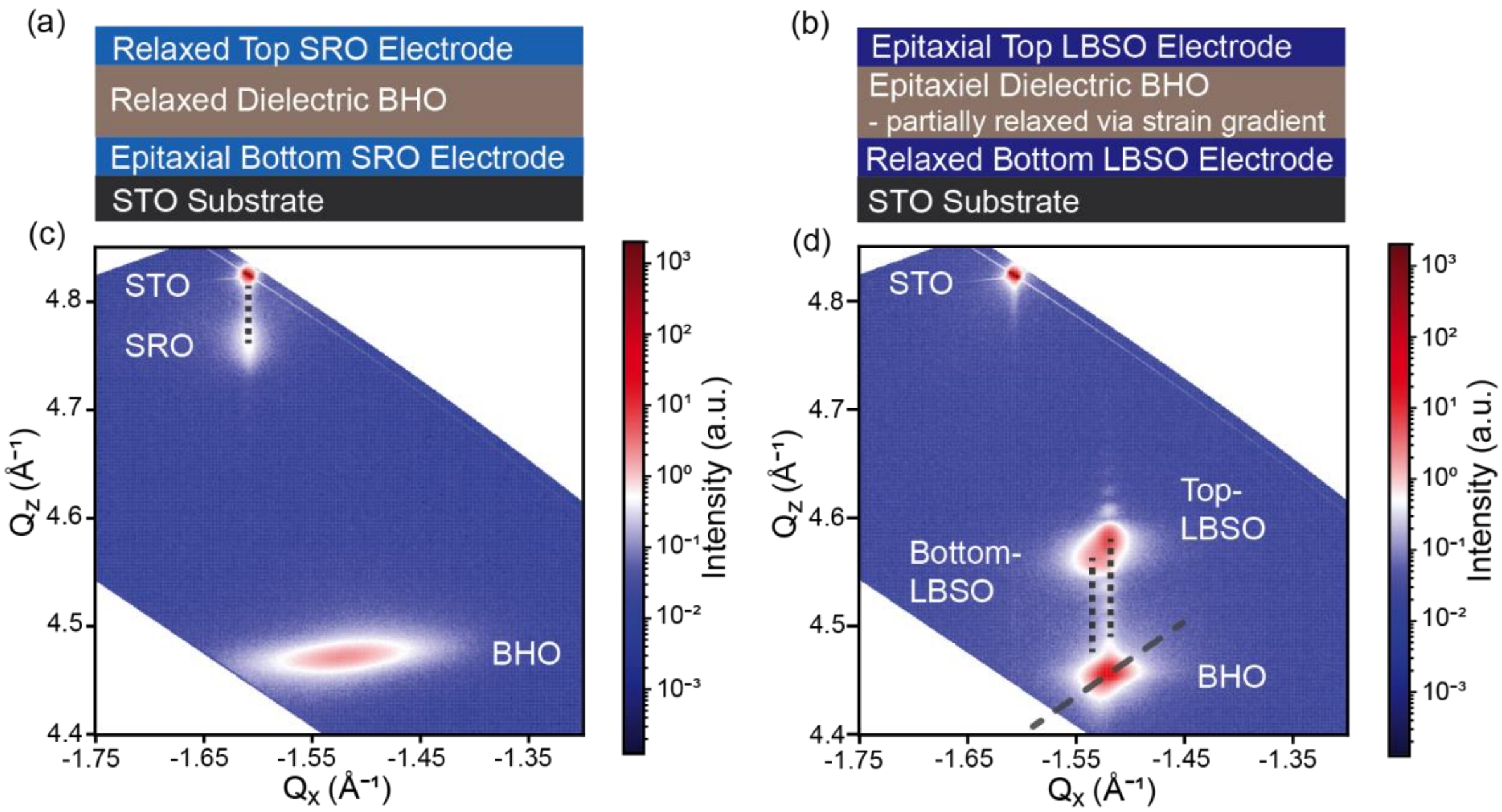

**Figure 2.** Schematic illustration of the **(a)** STO/SRO/BHO/SRO and **(b)** STO/LBSO/BHO/LBSO capacitor structure. The respective RSM's of the (-103) signal is shown in **(c)** and **(d)**. The vertical dotted lines correspond to the in-plane lattice parameters for two adjacent strained layers, whereas the diagonal line in **(d)** corresponds to the theoretical relaxation line of BHO. Lattice parameter for top and bottom LBSO electrode layers are distinguishable due to a relaxation gradient in the BHO signal.

STO, BHO and LBSO have a cubic crystal structure, whereas SRO (orthorhombic) is treated as pseudocubic. Thus, the films were anticipated to grow with a cube-on-cube epitaxial relationship. The RSM measurements were performed for the asymmetric (-103) signal to achieve an improved signal-to-noise ratio, due to the low incidence angle of the X-ray beam. **Figure 2 (c)** displays the RSM along the asymmetric (-103) Bragg signal of the capacitor structure with an SRO bottom and top electrode. SRO strains in-plane to STO (exhibiting the same $Q_x$ value), thus both peaks correspond to an in-plane lattice parameter of 3.91 Å. The SRO out-of-plane lattice parameter is 3.96 Å. The interposed BHO layer is fully relaxed according to the RSM and exhibits a lattice parameter of 4.15 Å in-plane and 4.22 Å out-of-plane. This position aligns with the fully relaxed BHO structure grown directly on the STO substrate **(Figure S1)**. The SRO top electrode is not visible in the RSM, indicating a reduced crystalline order due to growth on the fully relaxed BHO layer. These findings demonstrate that the dielectric BHO layer fully relaxes and does not accumulate residual strain when grown on SRO, due to the large lattice mismatch.

**Figure 2 (d)** shows the RSM of the capacitor structure with LBSO as both the bottom and top electrode. The fully relaxed bottom LBSO signal (in-plane 4.10 Å and out of plane 4.13 Å) is distinguishable from the top LBSO signal (in-plane 4.14 Å and out of plane 4.11 Å). The BHO layer strains to the bottom electrode and relaxes along the theoretical relaxation line (the dashed diagonal line overlaid on the BHO peak in Figure 2 (d)) based on the Poisson ratio of 0.25.[28] Such a relaxation profile corresponds to the elastic relaxation of the BHO.[30,31] The top LBSO electrode aligns with the partially relaxed BHO structure. The in-plane lattice parameter difference in the RSM between the bottom and top LBSO layers creates a 1% strain across the 100 nm BHO layer, corresponding to a strain gradient of approximately $10^6$ $m^{-1}$.

The high epitaxial quality of BHO is further verified by an HRXRD 2θ line-scan and the corresponding rocking curve **(Figure S2)**. Atomic force microscopy (AFM) and reflection high-energy electron diffraction (RHEED) measurements confirm the controlled layer-by-layer growth mode **(Figure S3)**, showing an atomically-flat film surface characterized by vicinal terraces that correspond to one unit cell in step height and a streaky profile exhibiting intensity oscillations during growth, respectively.

High angle annular dark field (HAADF) STEM analysis was performed to verify the strain relaxation mechanisms inferred from the RSMs**. Figure S4** includes a low-resolution cross-sectional image of the two prepared capacitor structures along with energy-dispersive X-ray spectroscopy (EDS) elemental maps. It should be noted that, for practical reasons, the second sample investigated by STEM and EDS employs an LBSO top electrode instead of a symmetric SRO/BHO/SRO configuration; however, this modification does not affect the analysis or interpretation of the results. Elemental maps confirm the uniform distribution of oxygen across both samples **(Figure S4 (a/b))**, while the distributions of cations indicate sharp transitions between the layers without extensive intermixing at the interfaces.

High-resolution STEM provides further insights into the relaxation mechanisms of the strained and relaxed capacitor structures (**Figure 3**). The lattice parameter differences among the bulk (unstrained) phases forming the STO/LBSO/BHO and STO/SRO/BHO systems are shown in **Figures 3(a)** and **3(b)**. **Figures 3 (c)** and **3 (d)** are overview images of both systems, which were analyzed for the presence of dislocations and changes in the in-plane lattice parameters with film thickness using geometric phase analysis (GPA, **Figure S5**). The analysis reveals that for the STO/LBSO/BHO system, LBSO relaxes through the formation of misfit dislocations that occur approximately every 8 nm at the STO/LBSO interface with 5.3% misfit (**Figure 3 (c)**), while the BHO film strains coherently to the underlying LBSO with very few dislocations observed at the interface in spite of the 2.1% misfit between the LBSO and BHO. In contrast, in the STO/SRO/BHO system, the SRO electrode grows coherently on the STO substrate due to the similar (pseudocubic) lattice spacings (only 0.6% misfit), and the BHO layer completely relaxes at the SRO/BHO interface through the formation of misfit dislocations approximately every 5 nm (**Figure 3 (d)**). In addition to the misfit dislocations at the SRO/BHO interface, also the BHO layer itself contains numerous defects (marked with arrows in **Fig. 3d**) and exhibits domains with slightly different orientation, as slight mistilt from the zone axis in high-resolution STEM images (mosaicity), which is consistent with its defect-mediated relaxation. Considering only dislocation density between the STO substrate and BHO functional layer, the STO/SRO/BHO system shows higher strain relaxation through the formation of misfit dislocations in comparison to the STO/LBSO/BHO system, implying that BHO is strained at the beginning of growth on LBSO.

The corresponding in-plane strain maps obtained using geometric phase analysis of the areas with low defect density inside the layers are shown in **Figures 3 (e)** and **3 (f)** for the strained and relaxed systems, respectively, along with the high-resolution images of the interfaces (**Figs. 3 (g-j)**) confirm these relaxation pathways. In the strained system, the LBSO/BHO interface is coherent and free of interfacial defects (**Figure 3 (g)**), whereas the STO/LBSO interface exhibits regular misfit dislocations (**Figure 3 (h)** In contrast, the relaxed reference system displays a defect-rich SRO/BHO interface (**Figure 3 (i)**), while the STO/SRO interface remains highly coherent (**Figure 3 (j)**).To quantify the strain distribution throughout both stacks, the in-plane ($\varepsilon_{xx}$) maps were analyzed quantitatively and the results are graphically presented in **Figure 3 (k)**. In the relaxed system, strain relaxation occurs abruptly at the SRO/BHO interface as evidenced by a sudden increase in the lattice parameter. The lattice parameter above the SRO/BHO are remains stable, some fluctuations are attributed to strain modulations in the vicinity of defects and/or domain boundaries distributed throughout the relaxed BHO layer. In the strained system, by contrast, strain relaxation is gradual - a sharp relaxation occurs only at the STO/LBSO interface (~11 nm), followed by continuous relaxation across the LBSO and BHO layers up to the end of the analyzed area at ~64 nm. Within the BHO layer, the strain changes for ~ 1% within the first ~ 26 nm of the BHO layer, which yields strain gradient of 3.9 • $10^5$ $m^{-1}$ (**Figure S5**). This value is about one order of magnitude lower than the strain gradient extracted from the RSM. However, this discrepancy can be explained when considering that the LBSO bottom electrode also exhibits a finite strain gradient across its thickness. If the gradual relaxation within LBSO is included, the total strain variation across the combined LBSO/BHO stack corresponds to roughly a 1% lattice parameter difference, consistent with the RSM-derived strain gradient.

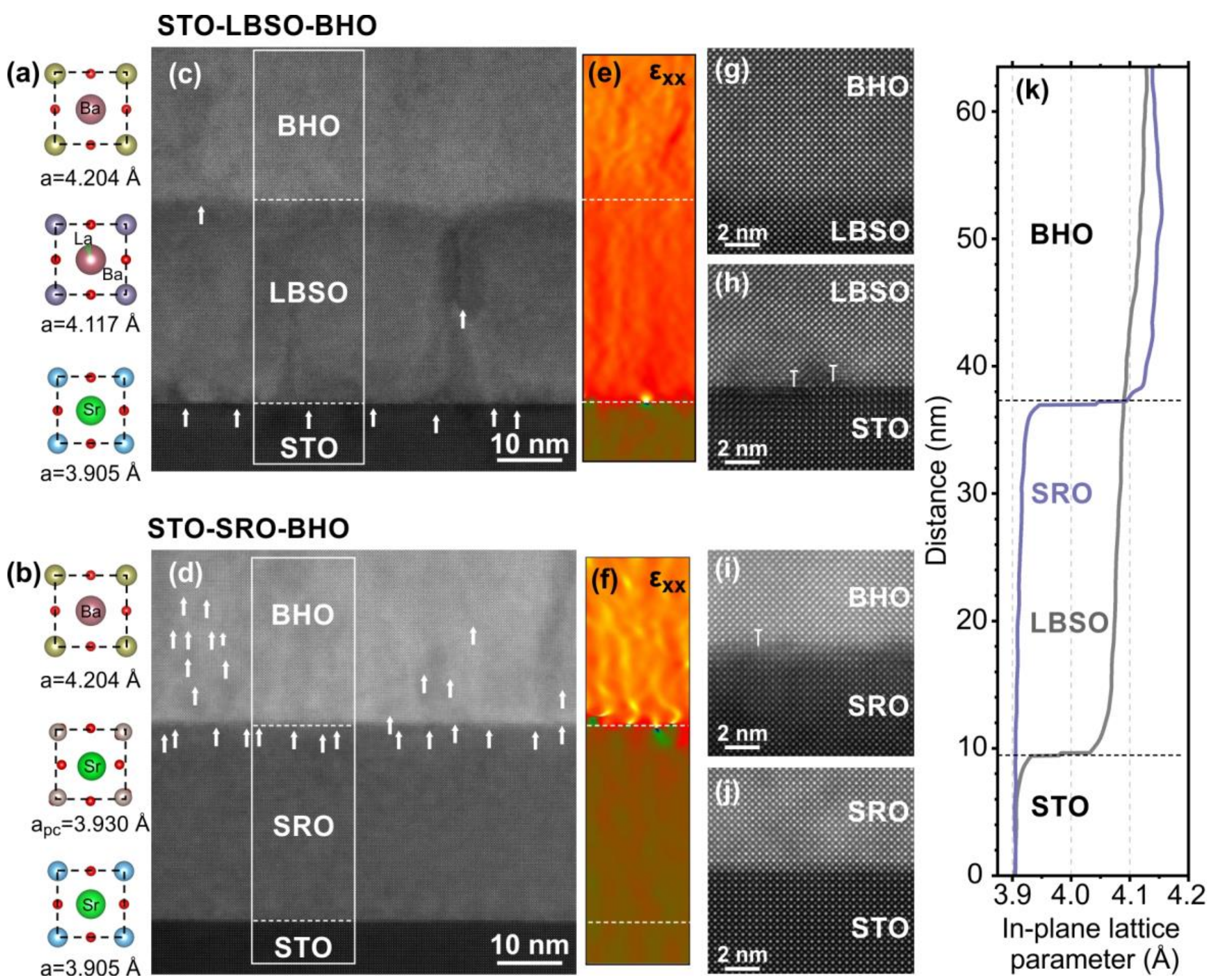

**Figure 3.** High resolution HAADF-STEM of **(top)** the STO/LBSO/BHO and **(bottom)** STO/SRO/BHO interfaces. **(a-b)** Elemental structures of each layer, **(c-d)** cross sectional STEM micrographs around the bottom electrodes**, (e-f)** corresponding strain maps, **(g-j)** high resolution interfaces, and **(k)** GPA analysis of the lattice parameter for the relaxed (purple) and strained (grey) systems as a function of distance along the thickness axis for the data in the boxes denoted in panels (c) and (d).

To confirm the absence of phase transitions in BHO, the relative permittivity (ε, **Figure 4 (a)**) and dielectric loss (tan(δ), **Figure 4 (b)**) were measured as a function of both temperature and frequency for the STO/LBSO/BHO/LBSO capacitor structure, in which the BHO layer is coherently strained to the LBSO and partially relaxed via a strain gradient. At frequencies below $2 \times 10^3$ Hz, increased noise levels are observed, as expected for low-frequency measurements.[32,33] Across the investigated temperature range from 50 K to 800 K, relative variations in the dielectric constant remain below 3.8%, and variations in dielectric loss remain below 3.0 %, indicating the absence of phase transitions in the BHO layer.

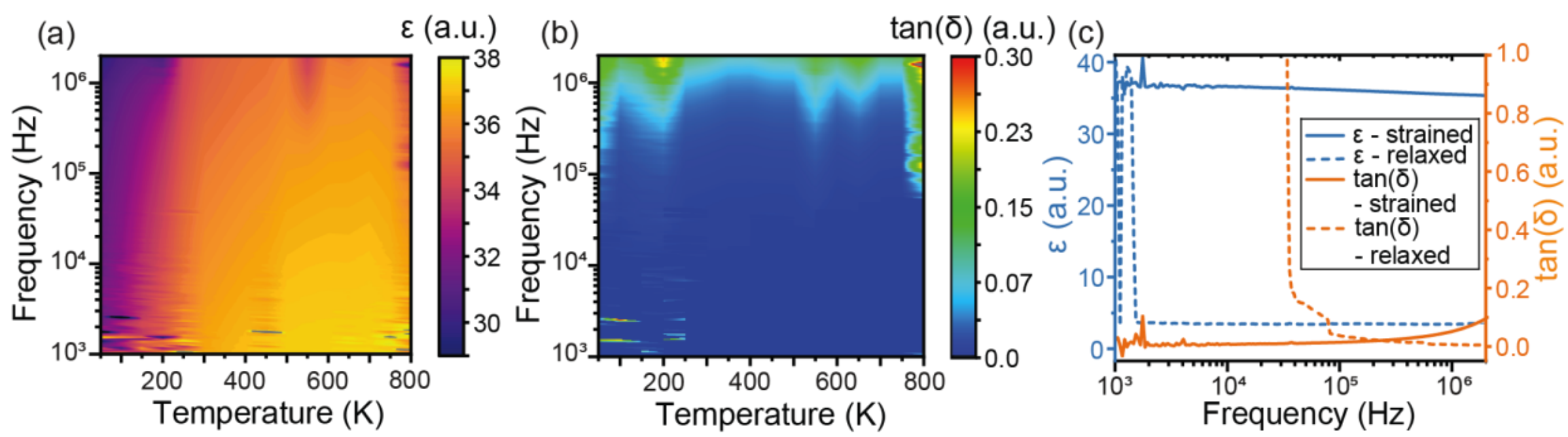

**Figure 4.** **(a)** Relative permittivity (ε) as a function of frequency and temperature. **(b)** Dielectric loss (tan(δ)) as a function of frequency and Temperature. **(c)** Relative permittivity (blue) and dielectric loss (orange) at 450 K for the fully relaxed BHO structure grown on SRO (dashed line) and BHO strained to LBSO, partially relaxed via a strain gradient (solid line).

Furthermore, a direct comparison was performed at 450 K between the STO/LBSO/BHO/LBSO and STO/SRO/BHO/SRO capacitor structures (**Figure 4 (c)**). For the structure with BHO strained to LBSO, the relative permittivity remains stable across the entire measured frequency range, with only a slight increase in dielectric loss near $10^6$ Hz. In contrast, the device incorporating a fully relaxed BHO layer grown on SRO exhibits significantly higher dielectric losses at low frequencies, which decrease as the frequency approaches $10^6$ Hz. This behavior suggests the presence of additional defect-mediated conduction pathways in the relaxed BHO layer. Moreover, the relative permittivity of the relaxed structure is more than four times lower than that of the strained system for frequencies above $2 \times 10^3$ Hz. These results indicate a substantial enhancement of the dielectric response in the strained BHO layer.

Notably, the dielectric constant of the fully relaxed system is more than four times lower than values reported in previous studies.[27,28] This deviation is attributed to an off-stoichiometry in the Ba:Hf ratio, with a deficiency for Ba corresponding to approximately 15%, as determined by Rutherford backscattering spectrometry (RBS) **(Figure S6)**. Such A-site cation deficiencies have been widely associated with pronounced alterations in the dielectric response of thin-film perovskites. Furthermore, this stoichiometric imbalance may account for the observed deviation from ideal tetragonality in the relaxed BHO (**Figure 1 (c), Figure S1**), which would otherwise be expected to equal one for a cubic perovskite material.[34-37]

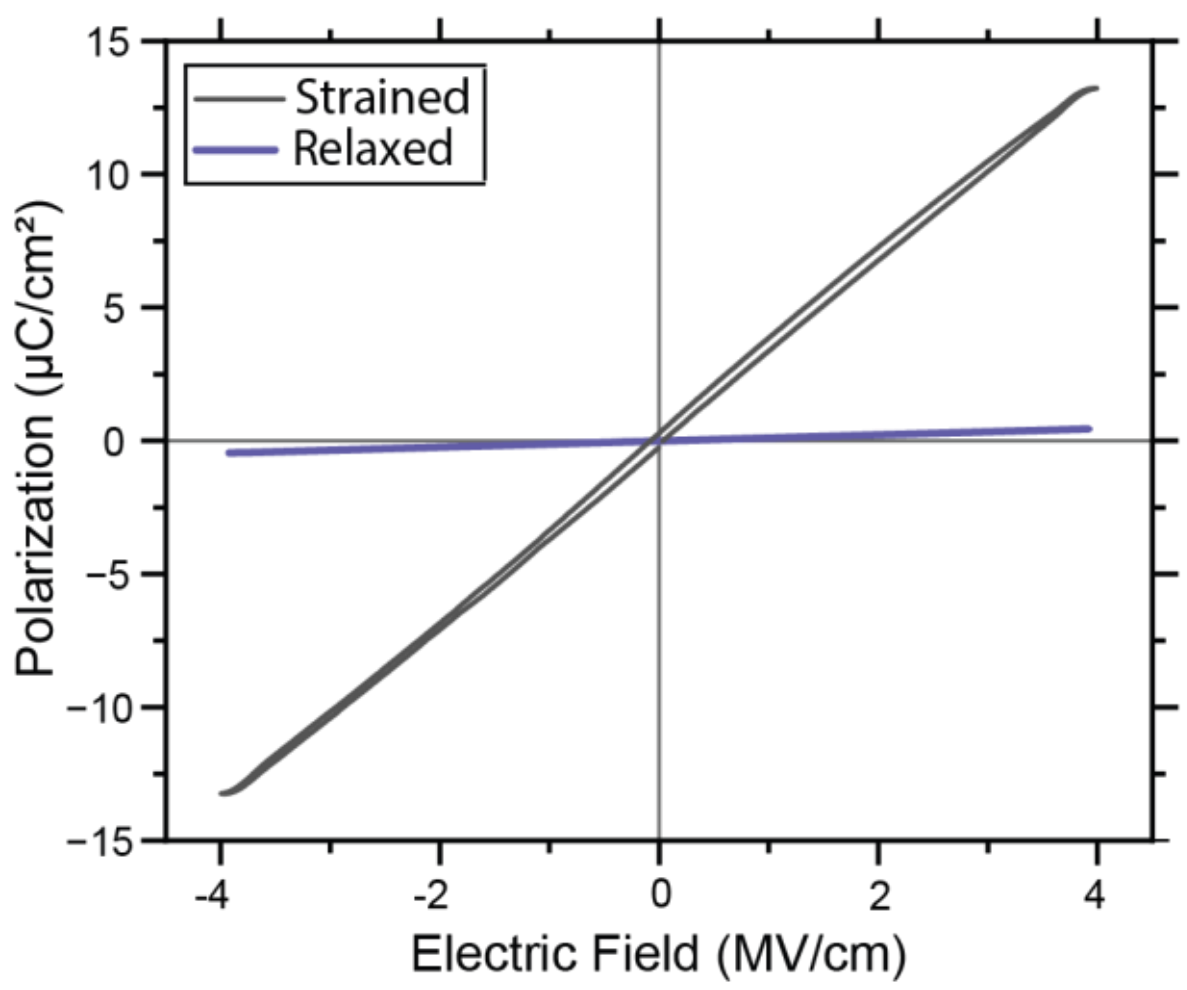

**Figure 5.** P-E hysteresis loop for the STO/SRO/BHO (fully relaxed)/SRO structure and the STO/LBSO/BHO/LBSO structure, with BHO strained to LBSO, partially relaxed via a strain gradient.

**Figure 5** presents the polarization-electric field (P-E) hysteresis loops for both capacitor structures. The STO/SRO/BHO/SRO device exhibits a maximum polarization of 0.45 μC $cm^{-2}$ at an applied field of 4 MV $cm^{-1}$, consistent with the expected linear dielectric behavior of a paraelectric material. In contrast, the STO/LBSO/BHO/LBSO structure, featuring a strain gradient in the BHO layer, displays a significantly enhanced polarization of 13.22 μC $cm^{-2}$ at the same electric field - a 29-fold increase compared to the unstrained structure. The P-E loop of the strained system shows a deviation from ideal linear paraelectric behavior, characterized by an enclosed hysteresis area and relaxation behavior near the maximum applied field. Additional measurements conducted at varying fields, frequencies, and temperatures (**Figure S7**) consistently confirm these trends. The STO/LBSO/BHO/LBSO device shows no signs of leakage up to an applied field of 4.5 MV $cm^{-1}$ (**Figure S7 (a)**) and a frequency as low as 10 kHz (**Figure S7 (b)**). The STO/SRO/BHO/SRO device exhibits a significant noise, due to its low maximum polarization (**Figure S7 (d,e)**). In both cases, the measured polarization remains constant between temperatures of 450 K and 650 K (**Figure S7 (c,f)**).

This study demonstrates that engineering strain gradients in centrosymmetric, paraelectric BHO thin films can significantly enhance polarization responses through flexoelectric coupling. By comparing two capacitor structures - one with BHO coherently strained on LBSO electrodes and another with BHO fully relaxed on SRO electrodes - this work isolates the influence of strain gradients on the polar behavior of the material.

RSM and TEM analyses reveal that BHO grown on LBSO exhibits coherent epitaxial strain and a continuous strain gradient across its thickness. In contrast, BHO on SRO relaxes *via* misfit dislocations, resulting in a defect-rich interface. This structural difference correlates with the observed electrical properties: the strained BHO on LBSO shows a 29-fold increase in polarization and a fourfold enhancement in relative permittivity compared to the relaxed BHO on SRO. This is the first time such a increase in polar response is achieved in a centrosymmetric material by introducing a strain gradient. To date, prior studies have focused on the polar response in either ferroelectric materials or systems exhibiting strain-induced phase transitions.[18,22,23,38,39]

The observed increase in relative permittivity in the strained BHO/LBSO structure contrasts with some prior studies on ferroelectric materials, which report a reduction in dielectric response in the presence of strain gradients.[18] However, both interfacial strain and defect density can significantly influence dielectric behavior. Interfacial defects are commonly associated with a suppression of permittivity due to scattering and disruption of lattice coherence, whereas coherent interfacial strain can enhance permittivity by altering the local polarizability and lattice dynamics.[40-45] Therefore, the enhanced polarizability observed here is likely the result of a complex interplay between the strain gradient, coherent interfacial strain, and the reduced defect density in the epitaxial BHO/LBSO system. Further experimental and theoretical work is necessary to decouple and quantitatively assess the individual contributions of these factors to the overall dielectric response.

The measured strain gradient of approximately $10^6$ $m^{-1}$ in the BHO/LBSO structure falls within the range known to produce significant flexoelectric polarization, as reported in previous studies on thin films.[1-3,18,22,23,46] However, the only centrosymmetric material studied is STO. Unlike STO, which can exhibit temperature and strain-dependent ferroelectric behavior, BHO remains paraelectric over a broad temperature range (50–800 K), in agreement with theoretical predictions.[28] The absence of any detectable phase transition in BHO throughout this range supports the conclusion that the observed polarization enhancement is attributable to strain gradient-induced flexoelectricity, not latent ferroelectricity or structural instabilities.[47-49] This is consistent with the understanding that flexoelectric effects can induce polarization in materials that are otherwise non-polar.[4,13-15]

Recent studies have emphasized the critical role of defects - such as vacancies, dislocations, and grain boundaries - in enhancing flexoelectric effects in dielectric materials. In particular, defect dipoles arising from either anion or cation vacancies have been shown to significantly increase the flexoelectric coefficient and flexocoupling response in linear dielectrics due to their reorientation under a flexoelectric field.[50,51] Similarly, strain gradients arising around dislocation cores or at grain boundaries have been linked to enhanced local flexoelectric responses and polarization shift currents.[52-54] Furthermore, Eliseev *et al.* [55] have introduced the concept of defect-driven flexochemical coupling, wherein inhomogeneous Vegard strains at defect-rich interfaces generate electric fields that strongly affect both polarizability and elasticity in thin films. The present work offers a comparative analysis of two capacitor structures: one featuring a controlled strain gradient, and the other undergoing strain relaxation through defect formation at the interface. Both structures have been grown from the same BHO target under identical deposition conditions, ensuring that compositional defects - such as Ba-site nonstoichiometry and oxygen vacancies - are approximately equal in the two samples. In the relaxed BHO/SRO structure, TEM reveals pronounced interfacial defects, while dielectric spectroscopy shows increased dielectric loss as well as a low polarization response in P-E measurements - suggesting that these defects do not enhance the macroscopic field-dependent polarization. In contrast, the coherently strained BHO/LBSO structure exhibits a sharp, dislocation-free interface, indicative of a high crystallographic quality, that results in a significantly enhanced polarization response. Thus, although defect-mediated flexoelectricity can play an important role at the local scale, this study highlights the significance of spatially extended, coherent strain gradients as a robust mechanism for enhancing macroscopic field-dependent polarization in centrosymmetric paraelectric thin films.

In this context, the possibility of domain or domain-wall-related contributions to the observed polarization needs to be carefully considered. However, several lines of evidence suggest that these mechanisms are not responsible for the enhanced hysteresis response; (1) BHO is a centrosymmetric, paraelectric perovskite with no ferroelectric phase transition, and our temperature-dependent dielectric measurements **(Figure 4)** show no anomalies that would indicate the emergence of polar domains; (2) the P–E hysteresis loops of the strained BHO/LBSO capacitor exhibit no imprint, back-switching, or frequency dependence typically associated with domain-wall dynamics (**Figure S6)**; [56, 57] (3) high-resolution cross-sectional TEM shows no signs of ferroelastic domains or structural modulations indicative of long-range polar order. Together, these observations limit the possibility of domain or domain-wall-mediated switching as the dominant source of polarization.

Overall, the strained BHO sample exhibits a 29-fold increase in polarization at an applied field of 4 MV/cm compared to the fully relaxed structure. To evaluate whether this enhancement can be attributed to flexoelectricity, we provide an order-of-magnitude estimate following Tagantsev's analytical relation (Equation 1). With a measured strain gradient of $\delta\varepsilon = 1\times10^{6}$ $m^{-1}$ and the observed polarization of 0.13 C $m^{-2}$, the corresponding flexoelectric coefficient corresponds to approximately$1.3\times10^{-7}$ C $m^{-1}$. This value lies within the range reported for perovskite thin films ($10^{-9}$ – $10^{-6}$ C $m^{-1}$), suggesting that flexoelectricity can reasonably account for the magnitude of the observed polarization.[1, 14,15,19] In addition, dielectric spectroscopy shows that the relative permittivity of the strained sample is four times larger than that of the relaxed sample, which contributes ~14% of the total polarization enhancement. [49] Therefore, the large enhancement is most plausibly associated with field-enhanced electro-mechanical coupling under strong strain gradients. [58–60] Additional contributions may arise from electrostriction, interfacial fields, local symmetry breaking, or incipient polar behavior.[59-62] At present, these explanations remain a hypotheses, and further dedicated experiments and theoretical modelling will be required to disentangle their relative importance.

## Conclusion

In summary, this work provides compelling evidence that strain gradient engineering via coherent epitaxial growth can effectively enhance the polar properties of centrosymmetric dielectrics through flexoelectric coupling. In this work, we have demonstrated the impact of strain gradient engineering on the polar properties of centrosymmetric, paraelectric BHO thin films by fabricating two capacitor structures *via* PLD: one incorporating a coherent BHO layer grown epitaxially on a LBSO bottom electrode (inducing a strain gradient), and another with a relaxed BHO layer grown on a mismatched SRO bottom electrode (without a significant strain gradient). RSM analysis confirms the presence of a substantial strain gradient (exceeding $10^{5}$ $m^{-1}$) in the BHO/LBSO heterostructure. Dielectric spectroscopy measurements revealed no phase transitions across a broad temperature range, confirming the intrinsic centrosymmetric and paraelectric nature of BHO. Remarkably, the strained BHO sample exhibited a 29-fold increase in polarization under an applied electric field of 4 MV $cm^{-1}$, alongside a significant enhancement in relative permittivity, compared to the relaxed counterpart. This provides the first direct experimental evidence of a strong correlation between an engineered strain gradient and enhanced polarization in a paraelectric perovskite oxide. Overall, this study establishes strain gradient engineering *via* coherent epitaxy as an efficient strategy for inducing and enhancing an electric polarization in non-polar materials through flexoelectric coupling. This approach offers a promising pathway for developing advanced dielectric materials and polar devices that do not rely on ferroic phase transitions.

## Experimental Section

### Pulsed Laser Deposition of the Thin Film Capacitor Structure

For each PLD process, a 10 × 10 × 0.5 $mm^2$ $SrTiO_3$ (100) substrate (Shinkosha) was used. The substrates were cleaned *via* sequential sonication: 10 min in acetone (Sigma-Aldrich), followed by 10 min in propan-2-ol (Sigma-Aldrich), and finally 30 min of chemical etching in a 1 M HCl solution in deionized water. After cleaning, the substrates were thermally annealed in a tube furnace (Heraeus AG) at 950 °C for 1 hour, with controlled heating and cooling rates of 6.2 °C $min^{-1}$ in an $O_2$ atmosphere. To ensure thermal contact during the PLD process, a 5 nm Ta adhesion layer followed by a 40 nm Pt layer were sputtered using a magnetron sputtering system (AJA) on the backside of the substrate.

Following preparation, the substrates were introduced into the PLD system (TSST) and heated at a rate of 10 °C $min^{-1}$ to the final deposition temperature. Thin films were fabricated using a pulsed KrF excimer laser (LPX305iCC, Lambda Physik) with a wavelength of 248 nm and a pulse length of 25 ns, ablating material from a disk-shaped target positioned 55 mm from the substrate with a spot size of 2.11 $mm^2$. All targets were purchased from Ultimate Materials Technology Co., Ltd.

The LBSO bottom and top electrodes were deposited at 720 °C with a laser fluence of 1.50 J $cm^{-2}$, an oxygen partial pressure of 5 × $10^{-2}$ mbar, a repetition rate of 2 Hz, and a total of 2000 pulses corresponding to a thickness of 40 nm. The SRO bottom and top electrodes were grown at 650 °C with an oxygen partial pressure of 6 × $10^{-2}$ mbar, a laser frequency of 1 Hz, a laser fluence of 2.00 J $cm^{-2}$ and 1800 pulses, corresponding to the same thickness as the LBSO. The BHO layer was deposited at 680 °C with a laser fluence of 1.75 J $cm^{-2}$, an oxygen partial pressure of 1 × $10^{-2}$ mbar, a repetition rate of 5 Hz, and a total of 10,000 pulses, corresponding to a thickness of 100 nm. After deposition, the temperature was reduced at a rate of 10 °C $min^{-1}$ to room temperature while maintaining the oxygen partial pressure of the top electrode. Overall, two different capacitor systems were grown: STO LBSO/BHO/LBSO and STO/SRO/ BHO/ SRO.

The growth was monitored *in situ* via a RHEED system (k-Space Associates, Inc.), operated at a emission current of 40 kV and an applied current of 1.5 A.

### Thin Film Morphology Characterization and Compositional Analysis

RSMs were performed using a Seifert 3003 PTS HRXRD equipped with a Cu Kα radiation source (λ = 1.5406 Å). The measurements were carried out using a line detector, which recorded intensity along 2θ with an angular resolution of 0.01°, and simultaneously employing an ω step width of 0.01°. The exposure time for each step scan was set to 140 seconds to ensure sufficient signal-to-noise ratio. The sample alignment was performed with respect to the (−103) reflection of the STO substrate.

AFM measurements were performed using a NanoSurf FlexAFM operating in tapping mode equipped with NCLR-50 cantilevers (Nano World). Images were acquired at a resolution of 512 × 512 pixels with a line scan time of 4 seconds.

Thin foil for STEM analyses was prepared with a focused ion beam (FIB; Helios Nanolab 650, Thermo Fisher Scientific, The Netherlands). The surface of the thin film sample was initially protected by 10 nm thin carbon layer to prevent surface charging. A 500 nm Pt layer was then deposited by electron beam-induced deposition (2 kV, 0.4 nA) followed by an additional Pt layer deposited by $Ga^+$ ion beam-induced deposition (30 kV, 0.23 nA), resulting in a final protective layer thickness of 2.5 µm. The lamellae were cut along the [100] direction of [001] oriented STO substrate. The in-plane direction corresponded to the electron beam direction (zone axis of the substrate) during STEM observations. Rough lamella chunks with dimensions of 12 × 8 µm were then milled perpendicular to the [001] surface normal, thinned to 2 µm, and transferred to a FIB grid using a micromanipulator. The lamella was first thinned to 200 nm using FIB at 30 kV by sequentially reducing the ion beam current from 780 pA to 80 pA. Finally, the lamella was sequentially polished on both sides using FIB at 5 kV (44 pA) and 2 kV (25 pA) until electron transparency.

STEM analyses were performed on a probe-corrected Spectra 300 (Thermo Fisher) microscope operated at 200 kV, equipped with high-brightness FEG (X-FEG) and Super-X EDS detector with four 30 $mm^2$ windowless, collimated SDDs for elemental mapping. High angle annular dark field (HAADF-STEM) images were recorded with a convergence angle of 21 mrad and detector collection angle of 57-200 mrad.

The film composition was analyzed by RBS using a 2 MeV $He^+$ ion beam at the Laboratory of Ion Beam Physics, ETH Zurich. The acquired spectra were evaluated using SIMNRA 7 simulations.

**Electrode Fabrication and Patterning**

The top electrode geometry was defined using a photolithographic process with a positive photoresist (S1813 G2, Kayaku Advanced Materials). The photoresist was spin-coated onto the sample at 3000 rpm for 60 seconds, followed by a soft bake at 114°C for 60 seconds to enhance adhesion and uniformity. A direct-write lithography system was used to define circular electrode patterns with a diameter of 75 µm, 50 µm and 25 µm, by selectively leaving these areas unexposed to UV radiation. The exposed photoresist regions were subsequently dissolved using a developer solution (AZ 726 MIF, Merck), leaving only the masked electrode areas protected.

After development, ion-beam etching (IBE) was used to remove the uncovered top layer material, defining the electrode structures, using either one of the following methods. Ar-ion beam etching was performed on patterned samples either with an Electrostatic Quadrupole Secondary Ion Mass Spectrometer (SIMS) (EQS-MS, Hiden Analytical Ltd.) or an Ionfab system (Oxford Instruments). In the former instrument, the sample was etched using a focused $Ar^+$ beam of ~100 µm diameter with an ion energy of 2.5 kV (-107 nA ion beam current) within an area of 2000×2000 $µm^2$. The selected spatial resolution was 400×400 pixels to keep a balance between etching speed and depth resolution. In order to distinguish the interface between LBSO and BHO, $La^+$, $Ba^+$ and $Hf^+$ were measured by the mass spectrometer during the etching process. An electron flood source (FS40, PREVAC) was used for charge compensation during the etching process. In the latter instrument, an 8 sccm Ar gas flow and an $Ar^+$ beam voltage of 250 V were used, corresponding to a beam current of 30 mA. During etching, the sample was rotated at 20 rpm and tilted at a 15° angle relative to the incident ion beam. The sample was etched for 10 min, corresponding to an etching depth of 40 nm (the thickness of the top electrode layer). The remaining photoresist was then removed using acetone. For the bottom contact, Ag paste (Sigma-Aldrich) was applied to ensure a reliable electrical connection.

## Dielectric Function and Polarization-Electric Field Hysteresis Measurements

The dielectric and P-E hysteresis measurements were performed using a cryogenic probe station (Advanced Research Systems) to ensure temperature stability and minimize external influences. The dielectric constant was measured using a Precision LCR Meter (E4980A, Keysight). A small AC voltage of 60 mV (corresponding to a field of 60 kV/cm) was applied to the sample, and measurements were conducted over a frequency range from $2 \times 10^2$ Hz to $2 \times 10^6$ Hz. The data acquisition was performed at 200 logarithmically spaced frequency points, with each point representing the average of five consecutive measurements to improve accuracy. For P-E hysteresis measurements, a Radiant Precision Multiferroic materials analyzer was used. The hysteresis loops were recorded with a measurement speed of 1 ms and a preset delay of 1000 ms to ensure charge stabilization. A total of 2000 measurement points were taken for each loop to accurately capture the polarization response.

## Acknowledgements

The authors wish to acknowledge funding from the Swiss National Science Foundation (SNSF) as part of an Ambizione fellowship (216196) and the project funding scheme (204103). ND acknowledges the financial support from the Slovenian Research Agency within research project J1-3025. We acknowledge the usage of the instrumentation provided by the Electron Microscopy Facility at PSI and we thank the EMF team for their help and support. JAB acknowledges funding from the European Union’s Horizon 2020 research and innovation programme under the Marie Skłodowska-Curie grant agreement No 884104 (PSI-FELLOW-III-3i). We thank the staff of the Laboratory for Nano and Quantum Technologies (LNQ) at PSI for cleanroom support.

## Conflict of Interest

The authors declare no conflict of interest.

## Data Availability

All relevant data is summarized in the Zenodo open data repository accessible via https://doi.org/10.5281/zenodo.17449344.

# Supporting Information

## Flexoelectric Polarization Enhancement in Paraelectric $BaHfO_3$ via Strain Gradient Engineering

*Timo Piecuch[1,2], Nina Daneu[3], Jeffrey A. Brock[1,4], Xiaochun Huang[1,2], Tina Radoševič[3], Arnold M. Müller[5], Christof Vockenhuber[5], Christof W. Schneider[1], Thomas Lippert[1,2], Nick A. Shepelin[1]**

[1]Center for Neutron and Muon Sciences, Paul Scherrer Institute, 5232 Villigen PSI, Switzerland

[2]Laboratory of Inorganic Chemistry, Department of Chemistry and Applied Biosciences, ETH Zürich, 8093 Zurich, Switzerland

[3]Advanced Materials Department, Jožef Stefan Institute, 1000 Ljubljana, Slovenia

[4]Laboratory for Mesoscopic Systems, Department of Materials, ETH Zürich, 8093 Zurich, Switzerland

[5]Laboratory of Ion Beam Physics, ETH Zürich, 8093, Zürich, Switzerland

## S1: Growth Quality

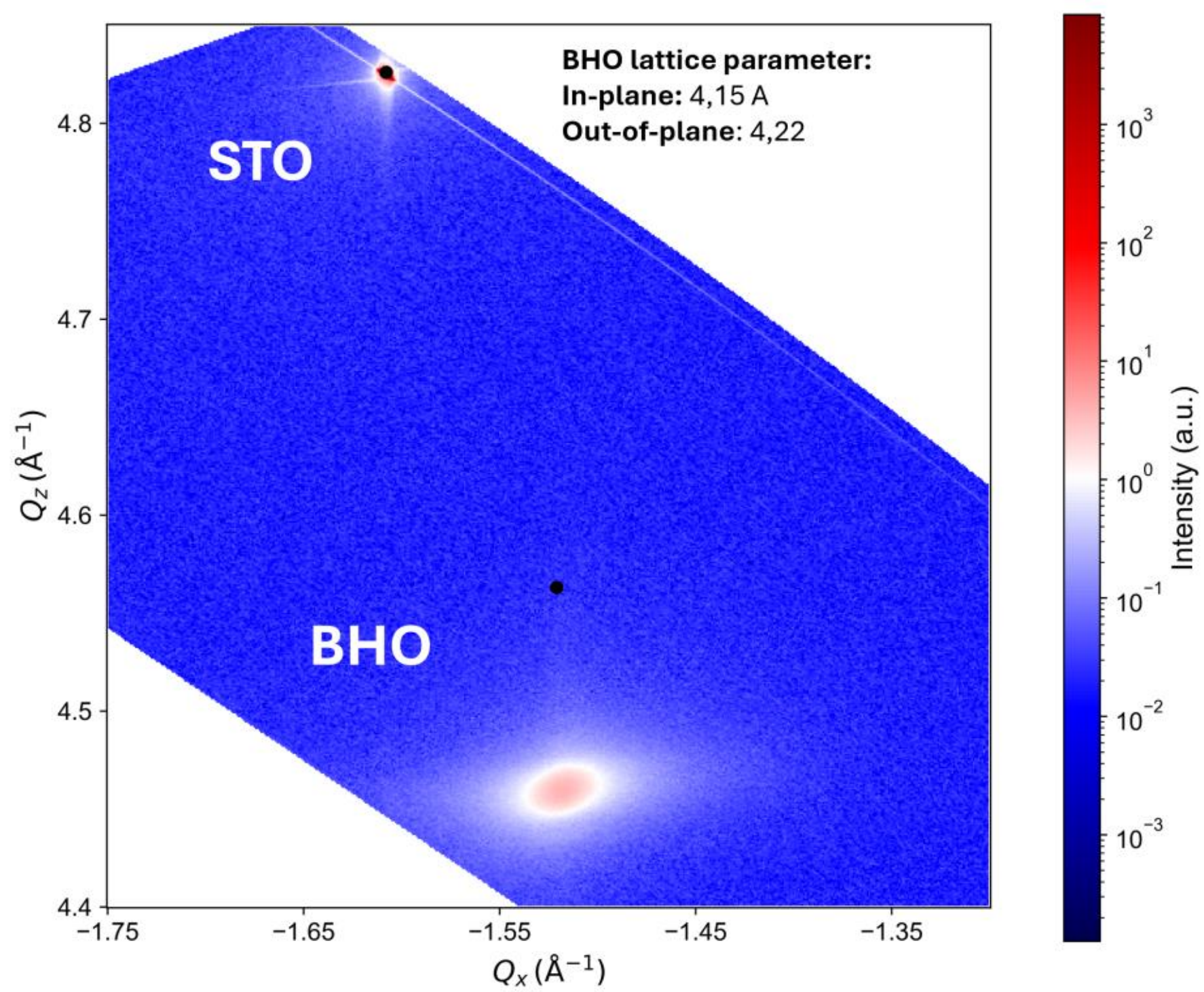

**Figure S1**. RSM of the asymmetrical (-103) signal of BHO directly grown on the STO substrate. The black dots indicate the literature positions of STO and BHO bulk materials. BHO is fully relaxed in-plane, due to a lattice mismatch of 6.8 % between substrate and film and the lack of overlap between the two peaks in $Q_x$. The in-plane and out-of-plane lattice parameters of BHO are indicated.

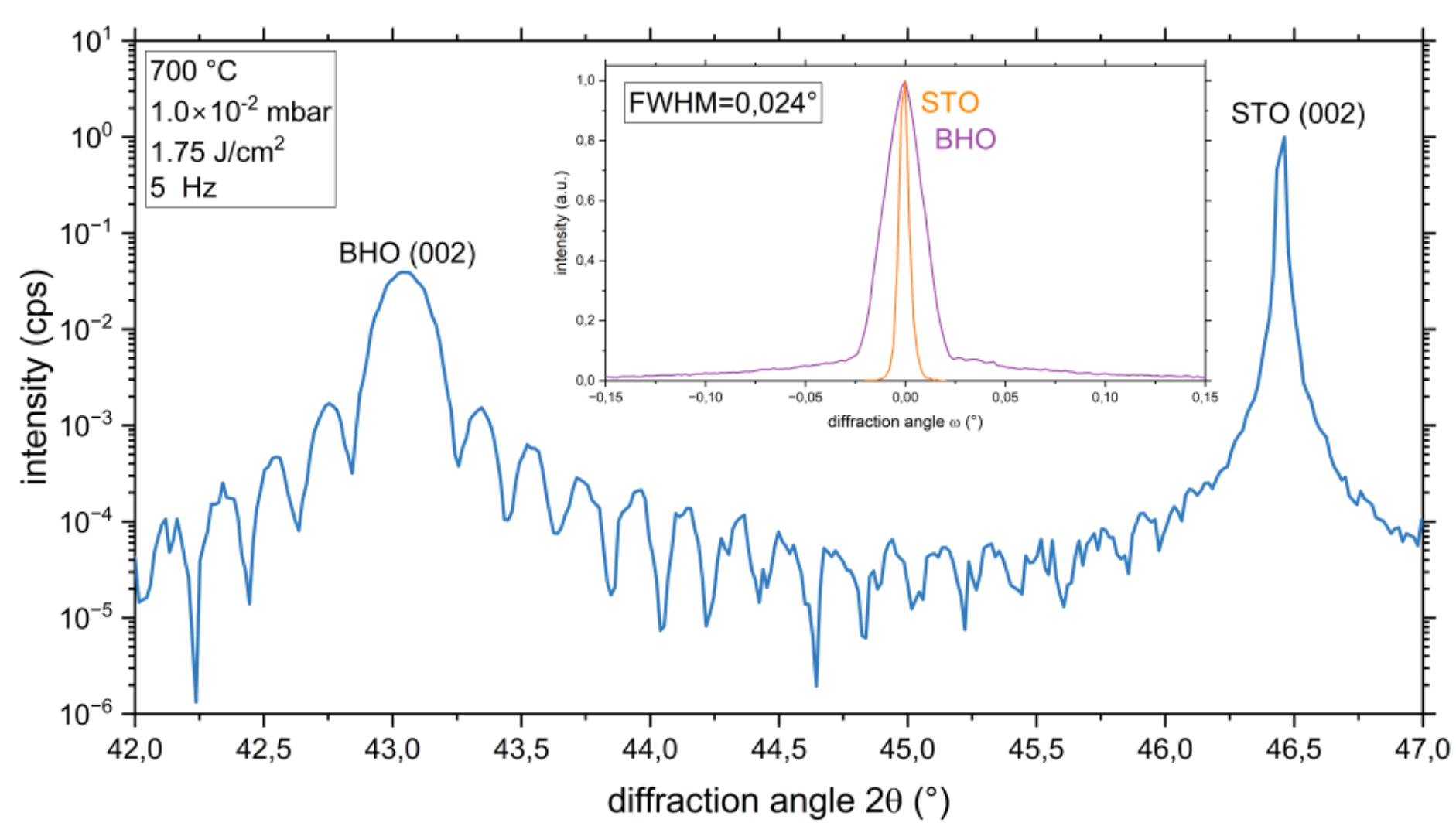

**Figure S2.** XRD 2θ line-scan of the 50 nm thick BHO grown on STO. The inset shows the rocking curve for the film and the substrate. Film full width at half maximum (FWHM) and growth parameters are indicated.

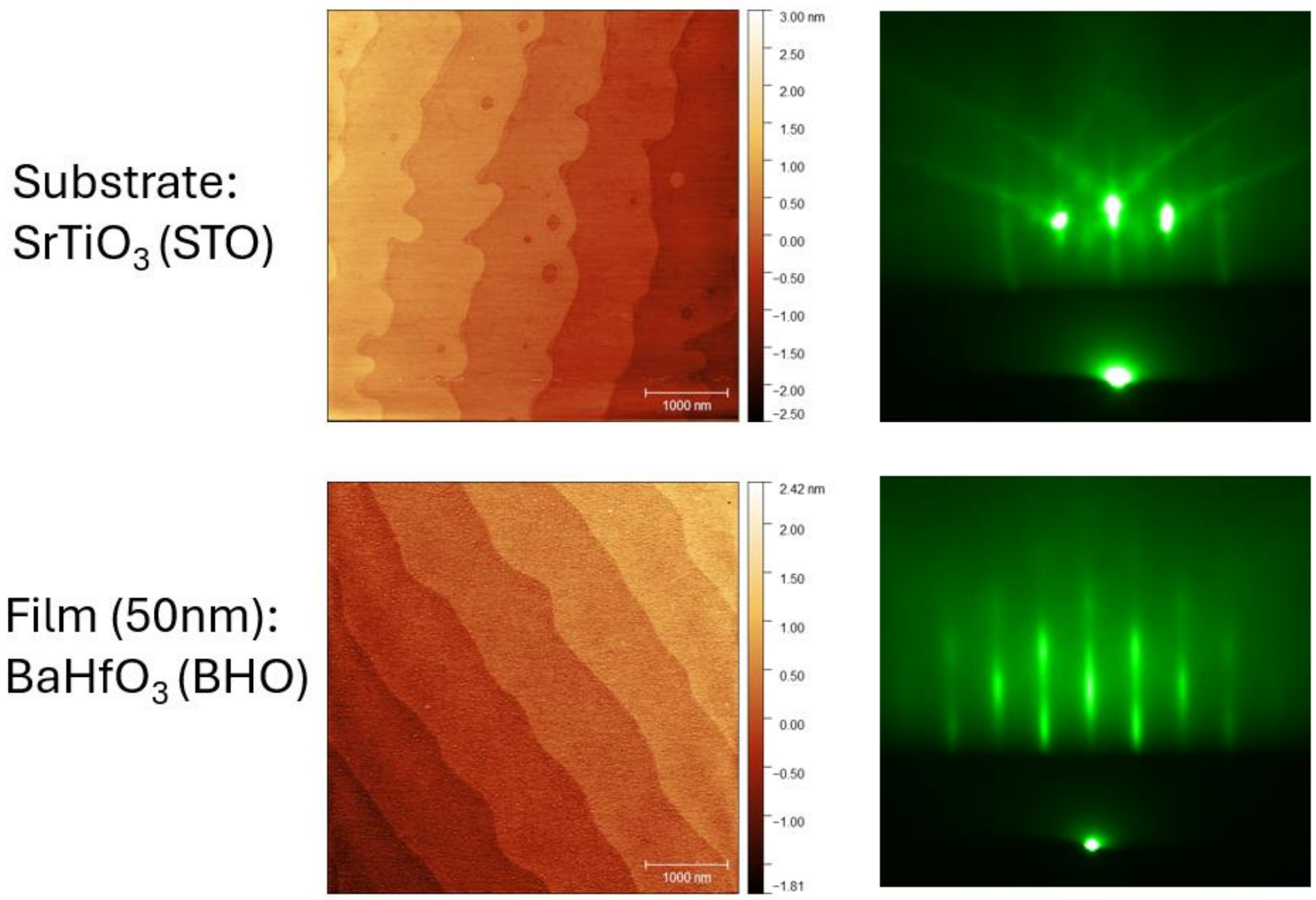

**Figure S3. (left)** AFM topography (5 × 5 μm) and **(right)** RHEED pattern before **(top)** and after **(bottom)** deposition of the 50 nm BHO film on the STO substrate employing the best growth conditions. The AFM data displays evenly spaced terrace steps, and the RHEED pattern shows streaks,

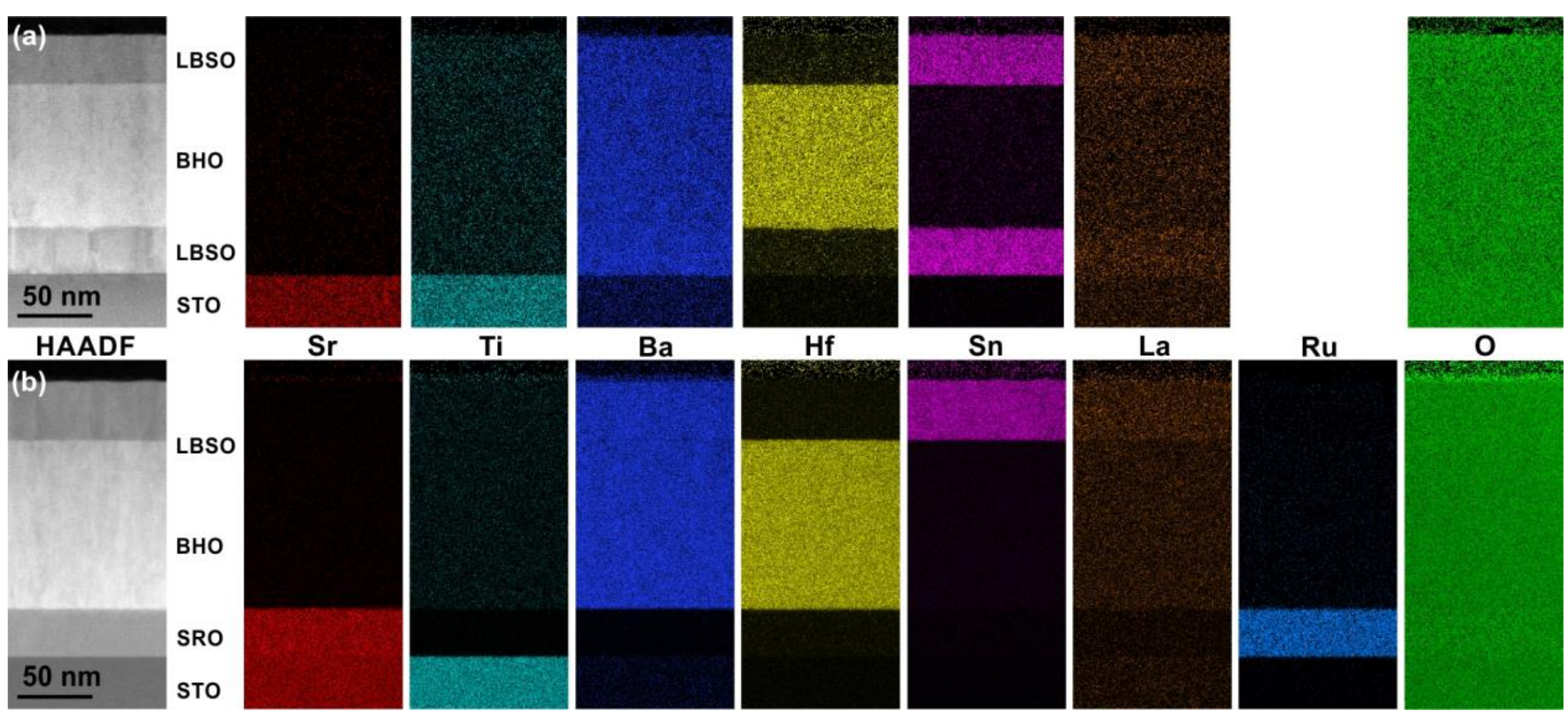

**Figure S4.** Cross-sectional TEM analysis of the capacitor structures: **(a)** STO/LBSO/BHO/LBSO and **(b)** STO/SRO/BHO/LBSO, including elemental mapping.

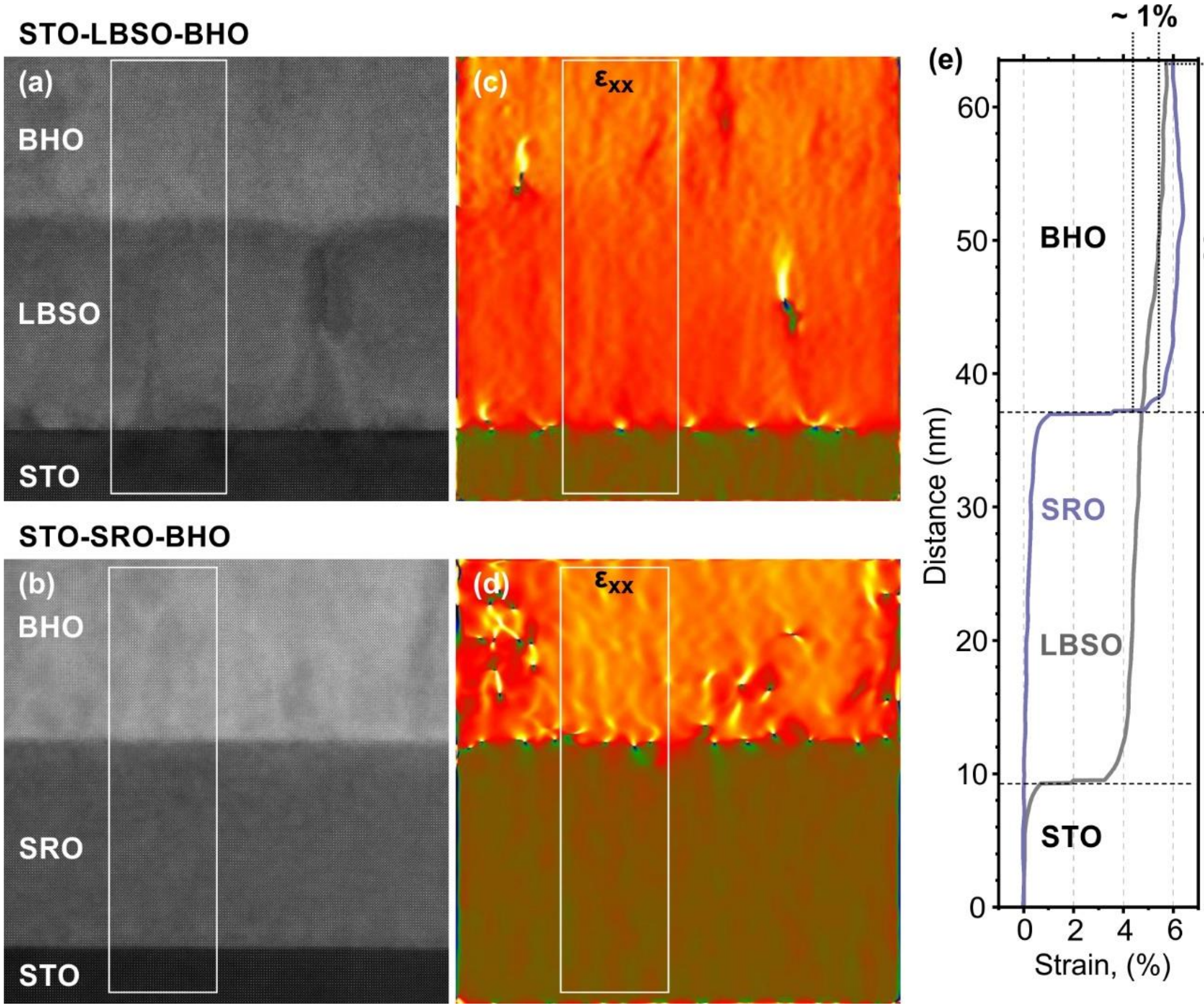

**Figure S5.** Atomic-resolution micrographs of the **(a)** STO/LSBO/BHO and **(b)** STO/SRO/BHO systems for geometric phase analysis (GPA). **(c,d)** GPA analysis reveals misfit dislocations at the different interfaces and also inside the layers. In addition, the maps were used for quantification of strain relaxation in the in-plane ($\varepsilon_{xx}$) direction. The graph in **(e)** shows that the strain in the STO/SRO/BHO system is completely released at the SRO/BHO interface, whereas gradual strain relaxation is observed in the STO/LSBO/BHO system from the STO/LBSO interface onwards. The strain gradient in approximately first 26 nm of the BHO functional layer was calculated based on the relative change of strain per thickness. In this case, strain relaxes for about 1% in the 26 nm, which yields strain gradient of about $3.9 \cdot 10^5$ $m^{-1}$.

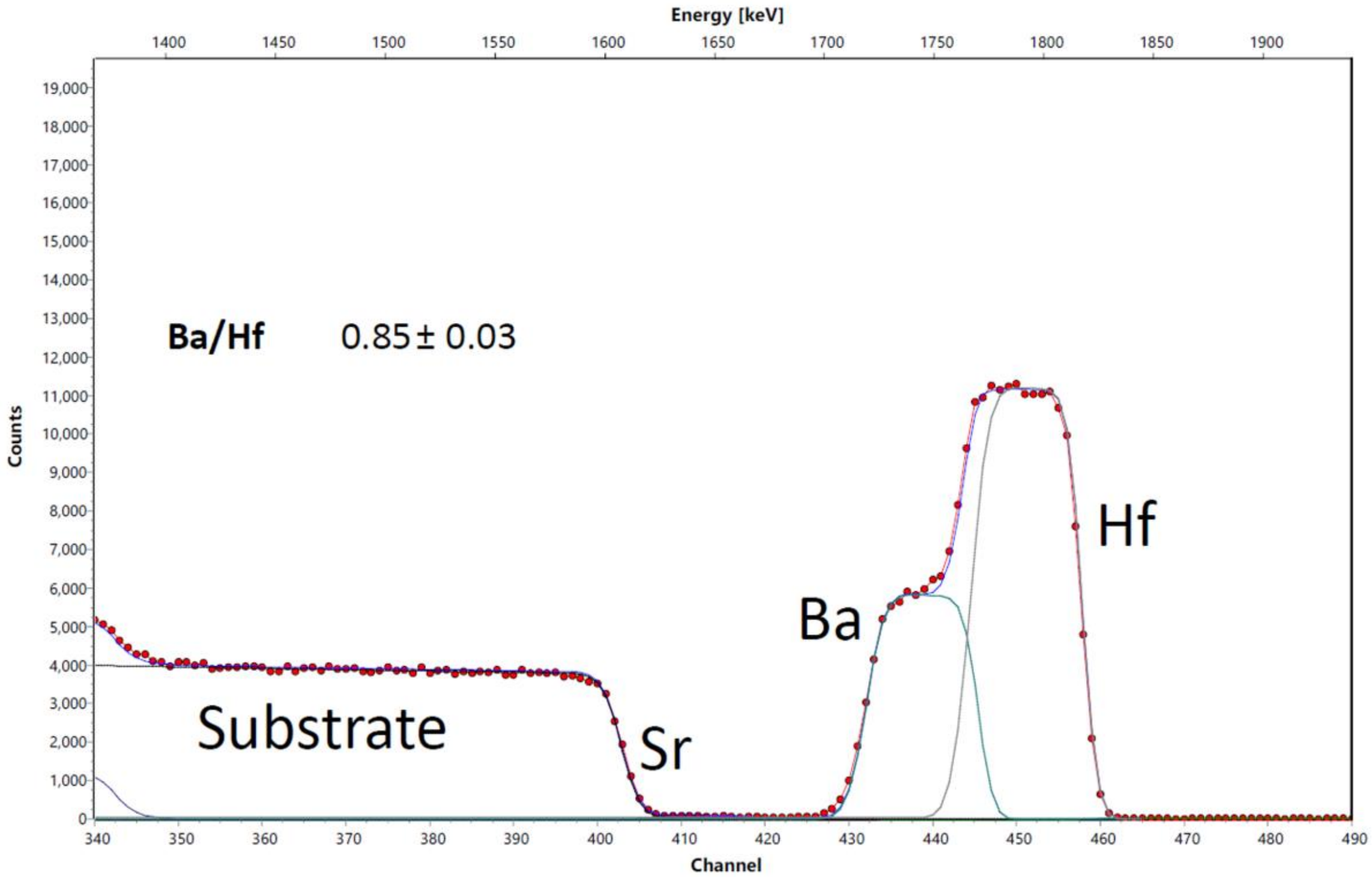

**Figure S6.** RBS spectra of the 50 nm BHO film grown on the STO substrate. Elemental cation ratio is indicated, exposing a Ba deficiency.

## S2: Electrical Properties

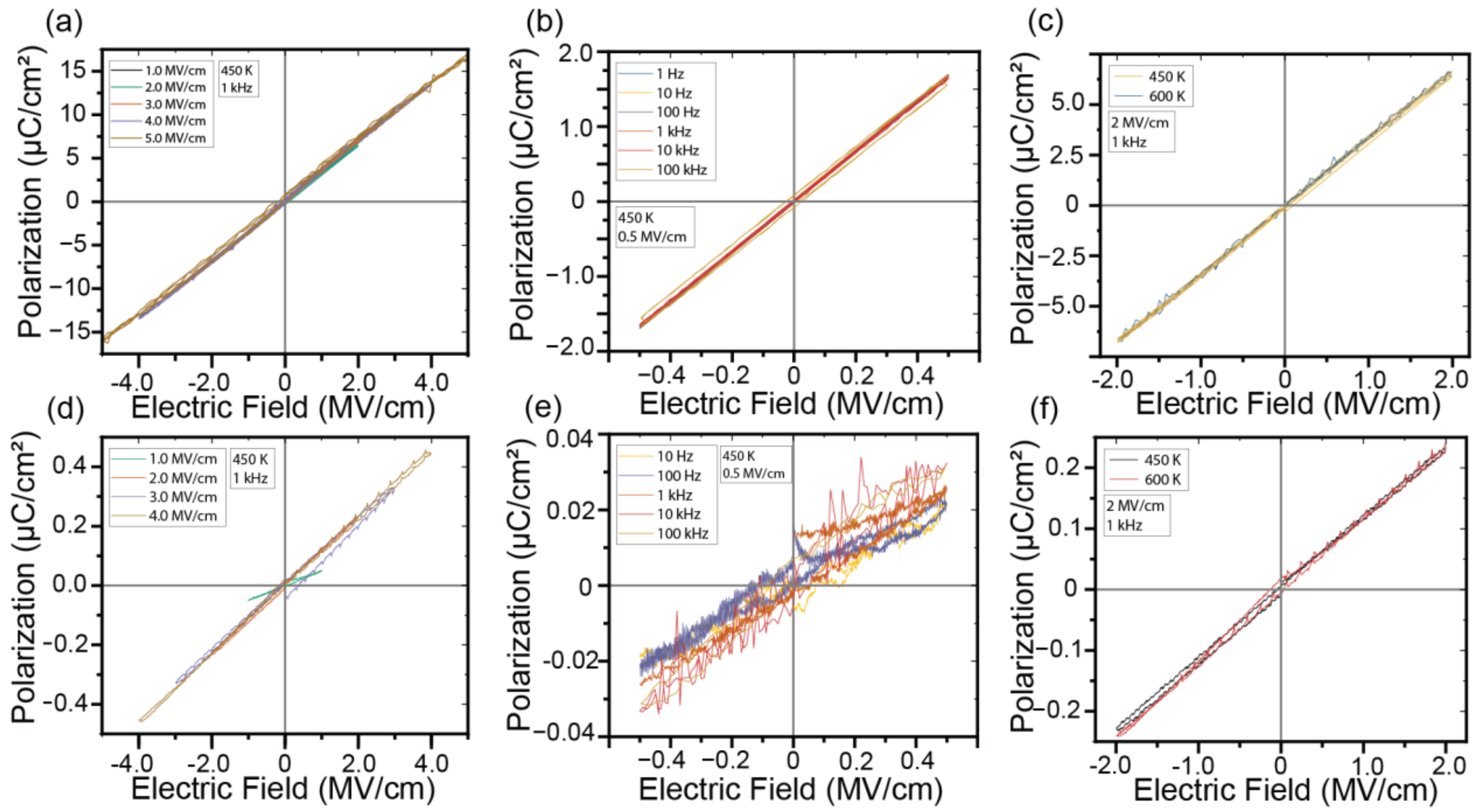

**Figure S7.** P-E hysteresis for the STO – SRO – BHO – SRO structure (**a-c**) and the STO – LBSO – BHO – LBSO structure **(d-f)**. The measurements have been performed for different fields **(a,d)**, different frequencies **(b,e)** and different temperatures **(c,f)**.